# Prediction of Ultraslow Magnetic Solitons via Plasmon-induced Transparency by Artificial Neural Networks

**Jiaxi Cheng,[1][*] Siliu Xu,[2][**] Shengwang Jiang,[3] and Zhiqiang Bo[4]**

[1]Department of Computer Science, University of Waikato, Hamilton, New Zealand
[2]School of Electronic and Information Engineering, Hubei University of Science and Technology, Xianning, China
[3]University of Electronic Science and Technology of China, Chengdu, China
[4] Software Engineering Institute, Xidian University, Xi'an, China
[*]Corresponding author: jc411@students.waikato.ac.nz  1601213918@sz.pku.edu.cn
[**]Corresponding author: xusiliu@hbust.edu.cn  xusiliu1968@163.com

**ABSTRACT**
Plasmon-induced transparency (PIT) in advanced materials has attracted extensive attention for both theoretical and applied physics. Here, we considered a scheme that can produce PIT and studied the characteristics of ultraslow low-power magnetic solitons. The PIT metamaterial is constructed as an array of unit cells that consist of two coupled varactor-loaded split-ring resonators. Simulations verified that ultraslow magnetic solitons can be generated in this type of metamaterial. To solve nonlinear equations, various types of numerical methods can be applied by virtue of exact solutions, which are always difficult to acquire. However, the initial conditions and propagation distance impact the ultimate results. In this article, an artificial neural network (ANN) was used as a supervised learning model to predict the evolution and final mathematical expressions through training based on samples with disparate initial conditions. Specifically, the influences of the number of hidden layers were discussed. Additionally, the learning results obtained by employing several training algorithms were analyzed and compared. Our research opens a route for employing machine learning algorithms to save time in both physical and engineering applications of Schrödinger-type systems.

**Keywords:** Magnetic soliton, Artificial neural network, Plasmon-induced transparency, Training algorithm

**1.0 Introduction**
There are many applications for machine learning, such as cancer prediction [1, 2], climate change [3, 4], and particle physics [5, 6]. Due to new insights in some fields, machine learning has significant impacts for not only scientists but also engineers [7, 8]. Artificial neural networks (ANNs) are a hot topic in machine learning and have been used as tools to perform research in optics and electronics during the past decade [9-11]. This work presents new experimental configurations and predictive models via ANNs.

Metamaterials are artificial electromagnetic media structured on subwavelength scales. Plasmon-induced transparency (PIT) is a phenomenon that is analogous to electromagnetically induced transparency (EIT) and is a productive way to study plasmonic polaritons in solid-state systems, which have much in common with EIT solitons in three-level cold atomic systems [12, 13]. PIT is a typical destructive interference effect that arises from the strong coupling between two types of modes of the meta-atoms in metamaterials. These two modes are dark modes and bright modes. The transparency window in the absorption spectrum occurs in both EIT and PIT, which has been observed in recent years [14, 15]. PIT metamaterials can be applied not only in electronic and optical devices [16-19] but also in materials [20, 21].

Recently, the simulation of the nonlinear Schrödinger equation (NLSE) in PIT has attracted extensive interest since new types of solitons have been predicted and observed [22]. However, in most cases, the exact solutions of a Schrödinger-type system cannot be acquired by analytical methods. Due to the relationships between the numerical results and the initial solutions and the parameters, the system design requires massive computing time to integrate the equations. This phenomenon creates a barrier to the application of numerical methods in our scheme.

In this article, we implemented machine learning algorithms by using neural networks to predict the outputs and propagation of solitons with different initial conditions, bypassing the approximate analytical and numerical PIT solutions. This paper is structured as follows. In section 2, the model is introduced. The simulation results for two types of magnetic solitons are shown in section 3, which demonstrates the evolution and peak power of the magnetic solitons, which depend upon the selection of two initial conditions. Section 4 compares the effects of the of the various types of input pulses on the predictions of the final expressions. In section 5, we discuss and compare the impacts of hidden layers with this model and algorithm, which is the same as that used in section 4. Some plots of the learning results for predicting the evolution of double solitons are shown in section 6. Section 7 gives the results of the effects of selected algorithms in neural networks, confirming that the LM method is the optimal algorithm in our model.

**2 Model**

The structure we used in this paper is a periodic array of unit cells, namely, magnetic meta-atoms with a pair of split-ring resonators (SRRs). In Figure 1(a), schematic drawings of SRR1 and SRR2 with inserted varactors are shown. The parameters are set as follows: the geometry of each resonator *l = 8 mm*, the thickness aluminum strips *w = 0.5 mm*, and the gap size of each resonator *h = 0.2 mm* to provide the nonlinear magnetic response [22]. Figure 1(c) shows the schematic structure of the metamaterial, including a periodic array of unit cells. For simplicity, the incident field is set along the z direction, and the electromagnetic field is set along the x and y directions. Note that the direction of the magnetic field is chosen along the y direction due to the necessary magnetic effects.

Figure 1(b) illustrates the RLC (Resistance, Inductance, Capacitance) circuit analog of the metamaterial. From this figure, it can be observed that several circuits represent two SRRs, which are bright oscillators and dark oscillators. In this case, the resonance frequencies of the SRRs are assumed to be the same for simplicity. In other words, $L_1 = L_2 = L$. The capacitance, resistance, and inductance of the RLC circuit are defined and shown in Figure 1(b) as C, R and L, respectively, for SRR1 and SRR2. Electromotive voltage is also contained in this model as the external excitation. From Kirchhoff's law, the relationships of the renormalized voltage *q* (see Ref. [23]) and time can be written as follows:

$$\ddot{q}_1 + \gamma_1 \dot{q}_1 + \omega_0^2 q_1 - \Omega^2 q_2 + \alpha_1 q_1^2 + \beta_1 q_1^3 = \omega_1^2 V(t) \qquad (1a)$$

$$\ddot{q}_2 + \gamma_2 \dot{q}_2 + (\omega_0 + \Delta)^2 q_2 - \Omega^2 q_1 + \alpha_2 q_2^2 + \beta_2 q_2^3 = 0 \qquad (1b)$$

In the above equations, $q_1$ and $q_2$ are the bright and dark oscillators, respectively. The damping rates of $q_1$ and $q_2$ are described as $\gamma_1$ and $\gamma_2$, respectively. The coupling efficiencies of the two resonators and the bright mode and external fields are defined as $\Omega^2$ and $\omega_1^2$, respectively. $\omega_0 \approx 2\pi \times 32$ *GHz* and $\omega_0 + \Delta$ are the linear resonance frequencies of RLC1 and RLC2, respectively. $\alpha_i$ and $\beta_i$ denote the second- and third-order nonlinearities, respectively, of the metamaterial. The detailed derivation of Equations (1a) and (1b) is shown in Ref. [23]. Figure 1(d) describes the structure of the neural network. The left side denotes the input layer, which consists of several inputs. A hidden layer is shown between the input layer, and the output layer on the right side. Pink, green and purple circular nodes represent the artificial neurons in the model. The arrows depict the connections between the output neuron and the input neurons. For deep learning, there are several layers, including input and output layers and hidden layers. Multiple connection patterns can be used between layers. Every connection has a weight whose value defines its importance. From the figure, it can be observed that every neuron possesses several connections that link the input and output neurons.

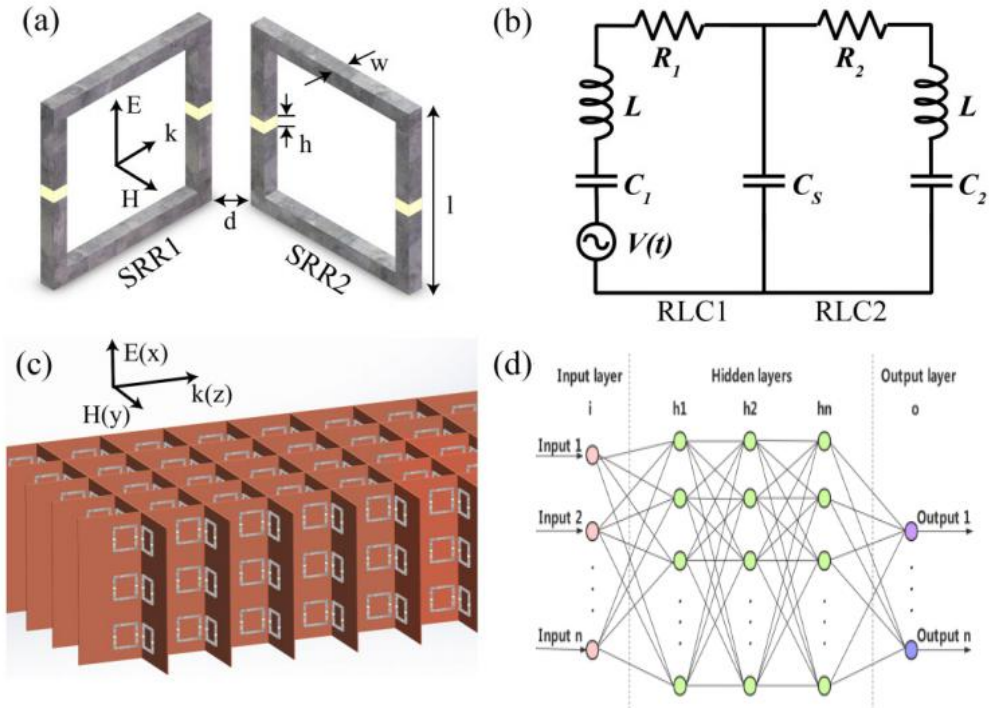

FIG. 1 (a) Metamaterial element consisting of two SRRs. The directions of the radiation field, magnetic field and electric field are mutually perpendicular. Bright and dark solitons are generated by selecting the propagation direction of incident radiation, which is connected to SRR1 and SRR2. The magnetic meta-atom parameters are *l*=8.0 *mm*, *w*=0.50 *mm* and *h*=0.20 *mm*. (b) Circuit analog of (a). RLC1 and RLC2 represent SRR1 and SRR2 in (a), respectively. (c) Schematic diagram of the PIT metamaterial consisting of a periodic array of metamaterial elements. The inlet describes the directions of k, E and H, which define the propagation, electric and magnetic fields, respectively.

## 3.0 Magnetic Solitons
In our system, the equation of motion for the magnetic field H is governed by Maxwell's equation:

$$\nabla^2 \vec{H} - \frac{1}{c^2}\frac{\partial^2 \vec{H}}{\partial t^2} = \frac{1}{c^2}\frac{\partial^2 \vec{M}}{\partial t^2} \qquad (2)$$

where the magnetic polarization intensity is given by $\vec{M} = \chi_D^{(1)}\vec{H} - N_0 S C_0 \frac{\partial q_1}{\partial t}\vec{e}$, where $\chi_D^{(1)}$ is the magnetic susceptibility of the linear substrate. The meta-atom density and the area of the split-ring resonator are defined by $N_0$ and S, respectively. Due to the direction of the radiation field of the dark oscillator, its magnetic moment contribution can be neglected, so $\vec{e}$ in Equation (2) represents the magnetic moment of SRR1.

After the second- and third-order nonlinear susceptibilities are acquired, we employ the method of multiple scales [24-27] and obtain a third-order approximation, which has the following form:

$$i\frac{\partial F}{\partial z_2} - \frac{1}{2}K_2 \frac{\partial^2 F}{\partial \tau_1^2} + \frac{c}{2\omega_0 n_D}(\frac{\partial^2}{\partial x_1^2} + \frac{\partial^2}{\partial y_1^2})F + \frac{\omega_0}{2cn_D}\chi^{(3)}|F|^2 F e^{-2\bar{\alpha}z_2} = 0 \qquad (3)$$

where $\tau_1$ is the generalized time parameter and $K_2$ is a coefficient that describes the group velocity dispersion. The envelope function F satisfies the divergence-free condition, $\frac{\partial F}{\partial z_1} + \frac{1}{V_g}\frac{\partial F}{\partial t_1} = 0$. The effect of linear absorption is described by the coefficient $\bar{\alpha} = \lambda^2 \operatorname{Im}(K)$, in which the absorption spectrum is Im(K). $\chi^{(2)}$ and $\chi^{(3)}$ are the second- and third-order nonlinear magnetic susceptibilities, respectively. In Equation (3), $n_D = (1+\chi_D^{(1)})^{1/2}$, and $x_1$, $y_1$ and $z_2$ are multiple variables. In our work, we assume that the diffraction effects based on the x and y directions are small enough to be neglected. In this case, Equation (3) can be modified to the following form:

$$i\frac{\partial U}{\partial s} + \frac{1}{2}\frac{\partial^2 U}{\partial \tau^2} + |U|^2 U = idU \qquad (4)$$

where s, $\tau$ are the generalized distance with z direction and time, respectively. Here, $d_0 = \frac{L_D}{L_A}$, where $L_D$ and $L_A$ represent the dispersion length and absorption length, respectively.

In this article, the pseudospectral method [28-29] is employed to solve the equation (3). Two initial cases are considered to study the ability of machine learning to predict the propagation and evolution of the magnetic solitons. Taking $d_0 = 1.2\times 10^{-3}$ [22], $u_0 = 1.2\times \operatorname{sech}[1.2(x+20)]\exp(ix)$ and $u_0 = 0.8\operatorname{sech}[0.8(x+20)]\exp(ix) + 1.2\operatorname{sech}(1.2x)$ as the initial conditions, we solve Equation (3) numerically. Figure 2 illustrates the evolution of the dimensionless radiation intensity $\frac{|u|^2}{|u_0|^2}$ with dimensionless time $\frac{t}{t_0}$ and distance $\frac{z}{L_D}$. Figure 2(a) presents the evolution of ultraslow magnetic solitons with the initial condition of a single pulse. It can be observed from the figure that there is no apparent deformation during the process of evolution. Figure 2(b) describes the propagation of two low-power solitons showing the collision of the two solitons. Two solitons can resume their shapes after they interact, which highlights their characteristic robustness.

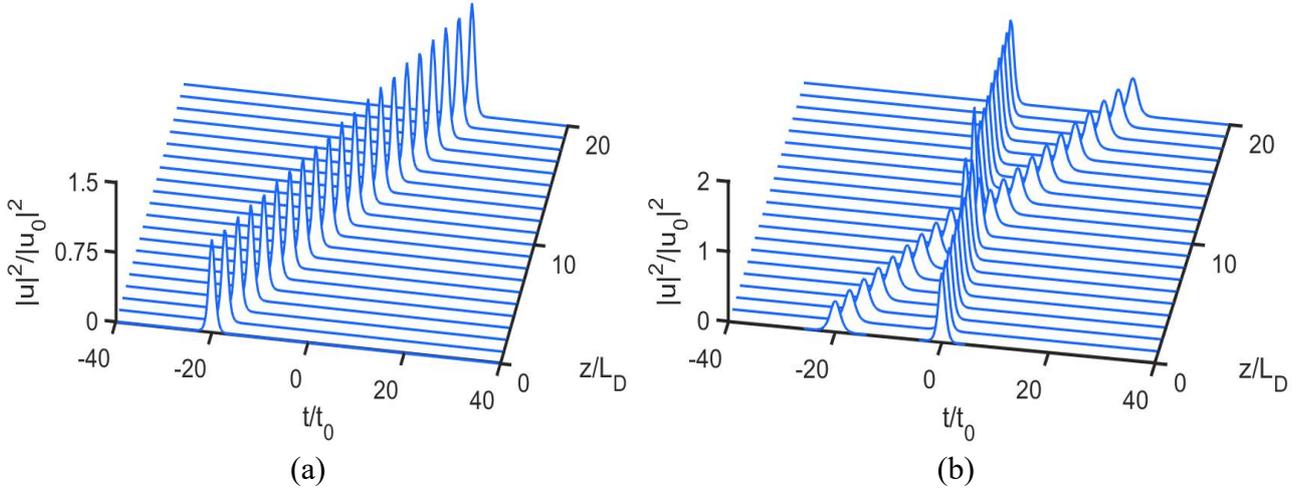

(a)   (b)

FIG. 2 Propagation of ultraslow magnetic solitons in the PIT metamaterial. (a) Evolution of a single input pulse $u_0 = 1.2 \times \text{sech}[1.2(x+20)]\exp(ix)$. (b) Evolution and interaction between two initial magnetic solitons $u_0 = 0.8\text{sech}[0.8(x+20)]\exp(ix) + 1.2\text{sech}(1.2x)$.

**4 Prediction of Magnetic Solitons**

In this section, the machine learning of single magnetic solitons and the collision of two ultraslow magnetic solitons will be discussed. In our model, the Levenberg-Marquardt (LM) algorithm [30] is employed as the training algorithm, and the performance is shown by the mean squared error (MSE), which is the average squared difference between the outputs and targets. The lower the MSE, the smaller the error. The algorithm usually needs more memory but less time. The process will terminate if the improvement stops, which can be guided by the values of the MSE for the validation sample data while training. In our scenario, 70% of the samples are used for training, 15% of the samples are for validation, and the remaining data are used for testing.

Figure 3(a1)-(a3) describes the learning output results for the single ultraslow low-power soliton with one hidden layer. The initial input pulse form is $u_0 = a\text{sech}[1.2(x+20)]\exp(ix)$, where a ranges from 1.0 to 1.5 with 1000 samples. Figure 3(a1) indicates the MSE versus the epochs. It is shown that the training stops after 14 epochs. The best validation performance is in the last epoch. Figure 3(a2) describes the error histogram versus instances with 20 bins defined by the number of vertical bars on the graph. This figure shows that the maximum probability of the error is approximately -0.01267. The regression value R estimates the relationship between outputs and targets. A value of 1 indicates a close correlation, whereas a value of 0 represents a random correlation. In Figure 3(a3), all R values are larger than 0.999, which means that the effects of training are perfect.

Figure 3(b1)-(b3) illustrates the effects of machine learning on the correlation of two magnetic solitons with four hidden layers. The initial condition is an input pulse with the form $u_0 = a\text{sech}[a(x+20)]\exp(ix) + \frac{3}{2}a \times \text{sech}(\frac{3}{2}ax)$, where a varies from 0.3 to 1.3 with 1000 values. The first figure shows the values of MSE versus the epochs. The more complicated relationship between the input data and the output data results in a larger number of epochs than that obtained for the single magnetic soliton. The best validation performance in this case is worse than the former case. Figure 3(b2) plots the error histogram, which shows that the range of error that contains the largest number of instances is greater than the case in the first row. From Figure 3(b3), we can see that the values of R are much greater, which indicates that the difficulties in predicting the output of the solitons vary with the initial conditions.

Figure 3(c1)-(c3) plots the training outputs of $\frac{z}{L_D}=20$ with 10 hidden layers for the initial pulse $u_0 = 1.2\text{sech}[1.2(x+b)]\exp(ix)$. The value of b varies from 0 to 20 with 1000 data points in this condition. Figure 3(c1) shows the MSE as a function of the epochs, and the best performance is 0.016991. As shown by the data plotted in Figure 3(c2), the performance of this error histogram is better than that in the second case. Four subplots of the values of R are shown in Figure 3(c3). The results reveal that it is more difficult to predict the final pulse when inputting a new value of b due to the translation of the solitons.

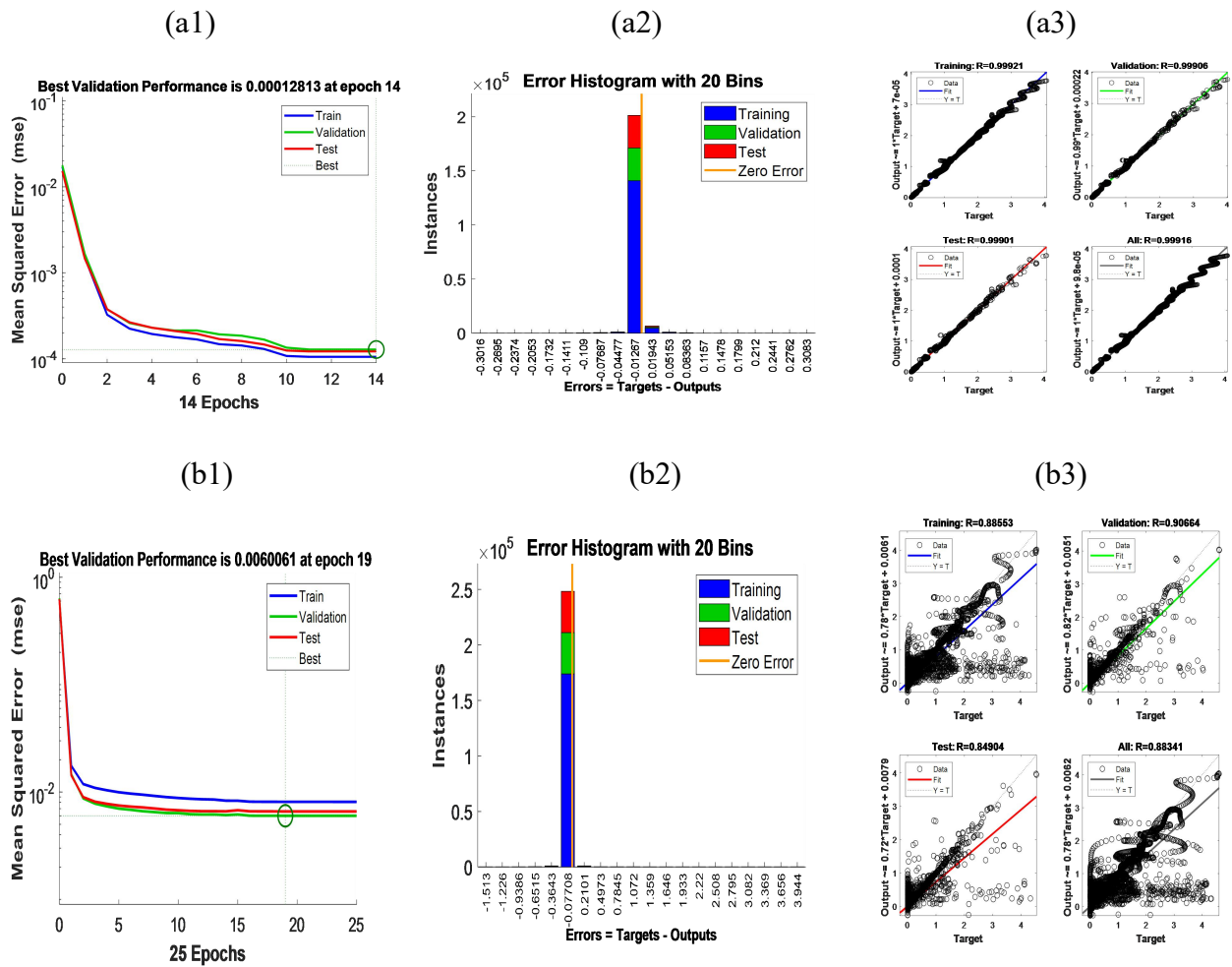

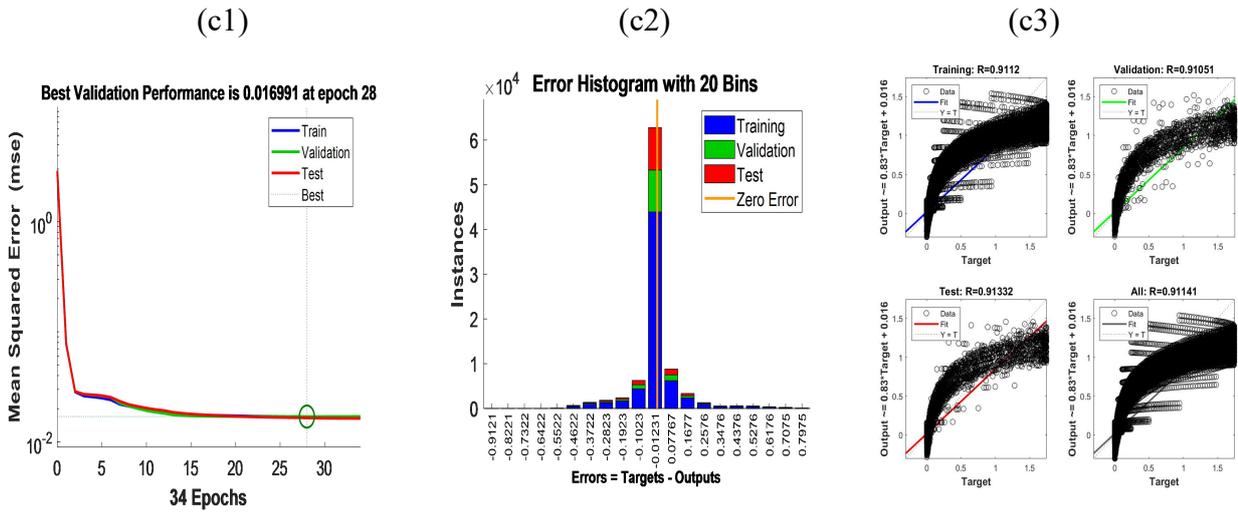

FIG. 3 Machine learning for predicting the outputs of single and two magnetic solitons by using the LM algorithm. (a1)-(a3) Learning the outputs of a single input magnetic low-power soliton with the predicted final shape of the soliton by changing the value of a in $u_0 = a\text{sech}[1.2(x+20)]\exp(ix)$. (b1)-(b3) Learning results for predicting the outputs of two magnetic low-power solitons by changing the value of a in $u_0 = a\text{sech}[a(x+20)]\exp(ix) + \frac{3}{2}a \times \text{sech}(\frac{3}{2}ax)$. (c1)-(c3) Training outputs of the predicted output of the solitons in the dimensionless distance $\frac{z}{L_D} = 20$ with varying values of the parameter b in $u_0 = 1.2\text{sech}[1.2(x+b)]\exp(ix)$. The first column shows the values of the MSE versus the epochs with the best performance. The second column represents the error histogram of instances. The last column plots the values of R for training, validation, testing and all data. The numbers of hidden layer(s) in the first, second and third rows are one, four and ten, respectively.

**5 The Effects of the Numbers of Layers**

In this section, the effects of choosing disparate numbers of layers will be explained. As discussed in Figure 1(d), the structure of a neural network contains input, output and hidden layers. Networks that have a larger number of hidden layers are more complicated in structure than networks with fewer hidden layers, which may result in better performance. The characteristics of multiple layers of the neural network will be discussed by an example of predicting the outputs of two magnetic ultraslow solitons.

Table 1 illustrates the predicted outputs for multiple layers in terms of b when inputting two interacting solitons. Note that in this case, the values of R in all instances do not apparently change for the last three conditions, whereas the training time is much longer than that in the first and second rows. Similarly, the best performance of the MSE does not significantly change in the last two cases. Considering all factors, the authors recommend the example that selects 3 hidden layers.

Table 1: Learning results of multiple layers to predict the outputs through inputting b of the collision between two magnetic solitons

| Number of hidden layers | Number of epochs | Computation time (taking the first case 1.000) | Best performance of MSE | R for all instances |
|---|---|---|---|---|
| 1 | 12 | 1.000 (46s) | 0.011758 | 0.81312 |
| 2 | 47 | 8.804 | 0.0081234 | 0.84979 |
| 3 | 32 | 11.065 | 0.0065969 | 0.87553 |
| 4 | 41 | 16.317 | 0.0060061 | 0.88341 |
| 5 | 22 | 15.652 | 0.0076577 | 0.88376 |

## 6 Predicted Evolution of the Solitons

We next extend our study to the second aspect of machine learning: the prediction of the evolution of solitons. Here, we focus our attention on double solitons as the initial solutions. In these cases, we select the hidden layers based on the consideration of all factors; we discuss this principle in the last chapter. The training approach is the LM algorithm, as in the former examples.

Figure 4 shows the steps to test and validate the learning efficiency of the neural network for the dynamics of dual magnetic solitons and to model their evolution. Figure 4(a) includes the MSE at all epochs, in which 0.010577 is the best result. The curves indicate that all types of data have the same variation trend. The MSE initially decreases rapidly; however, it is relatively stable for more than 5 epochs. Figure 4(b) presents the discrepancy between the model and the simulated data. The error bins show that the maximum error is -0.01967. Note that the model is still valid with the larger errors due to the tiny percentage of bins. The four dashed lines in Figure 4(c) present the targeted data, whereas the dashed lines define the training results for the relationship between outputs and targets. All values of R reach values larger than 0.91, which verifies the success of our model.

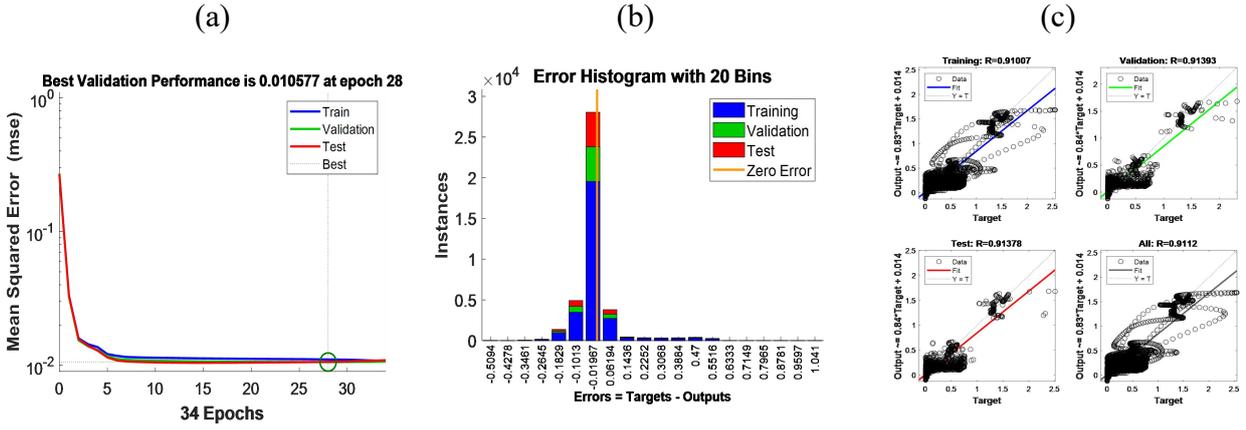

(a)　　　　　　　　　　　(b)　　　　　　　　　　　(c)

FIG. 4 Training results of the prediction of the evolution of double solitons by using the LM algorithm. (a)-(c) Double input magnetic low-power solitons with predicted evolution of the initial form $u_0 = 0.8\text{sech}[0.8(x+20)]\exp(ix) + 1.2\text{sech}(1.2x)$. Figure 4(a) plots the values of the MSE versus the epochs. Figure 4(b) is an error histogram of instances. The last figure shows the values of R in four cases. The number of hidden layers in the learning process is 4.

## 7 Discussion of the Algorithms

In addition to the LM algorithm, we compare and discuss other methods that can be employed. In the previous chapters, we explained how to predict the evolution and outputs of solitons by using the LM algorithm. In this section, we discuss two additional algorithms. The first method is Bayesian regularization (BR), which has been explained in the literature [31, 32]. This algorithm typically requires more time. However, it is always applied to the analysis of difficult, sparse and noisy data. The regularization method decides when to stop training. The second method is the scaled conjugate gradient (SCG) method [33. 34]. Less memory is required for this algorithm, and the process terminates if the generalization stops improving, which is indicated by the MSE of the validation instances.

These two methods are evaluated in machine learning by inputting the values of a in the form of an initial pulse $u_0 = a\text{sech}[a(x+20)]\exp(ix) + \frac{3}{2}a \times \text{sech}(\frac{3}{2}ax)$. Table 2 illustrates the comparison of different algorithms, including the Bayesian regularization and scaled conjugate gradient algorithms. The data in this table can be compared with the data in Table 2, in which the LM algorithm and several hidden layers have also been used to predict the outputs according to values of b for two magnetic solitons. As mentioned earlier, BR takes more time, but the performance of the algorithm is like that of SCG, which is indicated by the best-performing values of MSE and R for all samples. The optimal number of hidden layers for BR is 3 since the computation time is relatively short and R is higher than that for 4 hidden layers. It can be concluded from the table that as the number of hidden layers increases, the results undergo no apparent change. As a result, we can select the method with the shortest computation time. In contrast with Table 2, the efficiency of BR is better than that of the LM method, although the time is longer. The fastest speed of SCG proves the learning ability of this method, but the agreement between the simulated and predicted data is worse than that with the LM algorithm. The computation time of the LM method is between those of BR and SCG, which verifies the most appropriate algorithm under these conditions.

Table 2: Comparison of different algorithms

| Number of hidden layers/ algorithms | Number of epochs | Computation time (taking the first case of each algorithm 1.000) | Best performance of MSE | R for all instances |
|---|---|---|---|---|
| 2/ Bayesian regularization | 90 | 1.000 (835s) | 0.0098373 | 0.84972 |
| 3/Bayesian regularization | 83 | 1.581 | 0.0070884 | 0.89244 |
| 4/Bayesian regularization | 213 | 5.065 | 0.0075875 | 0.88716 |
| 2/scaled conjugate gradient | 147 | 1.000 (4s) | 0.012059 | 0.83259 |
| 3/scaled conjugate gradient | 117 | 0.5 | 0.011812 | 0.83601 |
| 4/scaled conjugate gradient | 149 | 1.5 | 0.0086142 | 0.83678 |

## 8 Conclusions

We have shown that machine learning can introduce new insights into the prediction of PIT. Specifically, we have verified that neural networks can learn not only single solitons but also double solitons in PIT. The learning abilities consist of predicting the outputs of the solitons and the evolution of the solitons in the temporal domain. To study the impacts of the hidden layers and algorithms, we create two tables to make comparisons and explain our results. Our results are especially significant for PIT, as it is the first application of machine learning in this area.

In terms of applications to physical systems, the authors expect that the neural network can be applied to induced transparency effects. Future applications may include more complicated neural networks to improve learning and model more complex physical systems. From a wider perspective, we expect that this work will benefit both experiments and simulations in optics, e.g., designing and analyzing optical experiments in EIT and PIT.

## Acknowledgements

The authors appreciate Dr. Richard Sanger of The University of Waikato for proofreading and discussions of this work. The authors also thank Mr. Zhenhao Cen for his kind help and debate.